\documentclass[pra,showpacs,showkeys,amsfonts,amsmath,twocolumn]{revtex4}
\usepackage{bm}
\newcommand{\si}[1]{\sigma_{#1}}

\newcommand{\ro}{\rho}

\newcommand{\R}{\mathbb R}

\newcommand{\tr}{\mathrm{tr}\,}

\newcommand{\tl}[1]{\boldsymbol #1}

\begin{document}
\title{Bell inequalities and linear entropy.
Comment on the paper of E. Santos :\\
"Entropy inequalities and Bell inequalities for two-qubit systems"}

\author{L. Jak{\'o}bczyk}

\affiliation{Institute of Theoretical Physics, University of
Wroc{\l}aw,
Pl. M. Borna 9, 50-204 Wroc{\l}aw, Poland}

\begin{abstract}
It is shown that even if the
linear entropy of mixed two-qubit state is not smaller then
0.457, Bell - CHSH inequalities can be violated. This contradics
the result obtained in the paper of E. Santos \cite{Santos}.
\end{abstract}

\maketitle

As is well known, the  relation between violation  of Bell
inequalities and entanglement properties of quantum states is not
clear. Although all pure entangled states violate Bell
inequalities, for mixed states entanglement is not equivalent to
such violation. One has to consider also the mixedness of quantum
state $\ro$, measured for example by their linear entropy
\begin{equation}
S_{12}(\ro)= 1-\tr \ro^{2}
\end{equation}
In  Ref. \cite{Santos}, the author claims that in a two-qubit
system, states $\ro$ for which
\begin{equation}
S_{12}(\ro)\geq \frac{1}{\sqrt{2}}-\frac{1}{4}\simeq 0.457
\end{equation}
satisfy all Bell - CHSH inequalities. This is however
inconsistent with the results obtained in \cite{DJ}. In that paper
we have investigated the relationship between entanglement
measured by concurrence $C$, mixedness measured by normalized
linear entropy $S_{L}=\frac{4}{3}S_{12}$ and CHSH violation, for
a class of mixed two-qubit states. One of our results shows that
for suitable values of $S_{L}$ and $C$, we can find states
$\ro_{1}$ and $\ro_{2}$ with equal linear entropies and
entanglement, such that $\ro_{1}$ violates CHSH inequalities
whereas $\ro_{2}$ fulfills these inequalities. It appears that
condition (2) still admits the existence of such states. Consider
for example the states
$$
\ro_{1}=\begin{pmatrix} 0&0&0&0\\[2mm]
0&0.549027&0.125&0\\[2mm]
0&0.125&0.449798&0\\[2mm]
0&0&0&0.001175
\end{pmatrix}
$$
and \vskip 4mm\noindent
$$
\ro_{2}=\begin{pmatrix} 0&0&0&0\\[2mm]
0&0.632864&0.125&0\\[2mm]
0&0.125&0.317431&0\\[2mm]
0&0&0&0.049708
\end{pmatrix}
$$
By a direct calculation, one can check that
$S_{12}(\ro_{1})=S_{12}(\ro_{2})=0.465$, so inequality (2) is
satisfied, but
\begin{equation}
\max\limits_{B}|\tr (\ro_{1}B)|=2.05699
\end{equation}
and
\begin{equation}
\max\limits_{B}|\tr (\ro_{2}B)|=1.86929
\end{equation}
Here
$$
B=\tl{a}\cdot\tl{\sigma}\otimes
(\tl{b}+\tl{b}^{\prime})\cdot\tl{\sigma}+\tl{a}^{\prime}\cdot\tl{\sigma}\otimes
(\tl{b}-\tl{b}^{\prime})\cdot\tl{\sigma}
$$
with $\tl{a},\,\tl{a}^{\prime},\,\tl{b},\,\tl{b}^{\prime}$  unit
vectors in $\R^{3}$ and $\tl{\sigma}=(\si{1},\,\si{2},\,\si{3})$.
To obtain equalities (3) and (4), one can use the important
result of Ref. \cite{HHH}:
$$
\max\limits_{B}|\tr (\ro B)|=2\sqrt{m(\ro)}
$$
where $m(\ro)=\max\limits_{j<k} (u_{j}+u_{k}),\; u_{j},\;
j=1,2,3$ are the eigenvalues of the matrix $T_{\ro}^{T}T_{\ro},\;
T_{\ro}=(t_{nm}),\; t_{nm}=\tr (\ro\, \si{n}\otimes \si{m})$.
\par
We see that $\ro_{1}$ violates CHSH inequalities and $\ro_{2}$
fulfills these inequalities, despite of the fact that both of them
satisfy inequality (2). Actualy, we are able to construct one
parameter family of states satisfying (2) and violating CHSH
inequalities.
\par
So we come to a conclusion that the result stated in Theorem 1 of
\cite{Santos}, cannot be true.

\end{document}